# Compact Ka-Band Metalens Antenna Enabled by Physics-Assisted Particle Swarm Optimization (PA-PSO) Algorithm

*Shibin Jiang, Wenjun Deng and Weiming Zhu\**

Shibin Jiang, Wenjun Deng and Weiming Zhu

School of Optoelectronic Science and Engineering University of Electronic Science and Technology, Chengdu 610051, China

E-mail: zhuweiming@uestc.edu.cn

**Abstract**

The design of multiple-feed lens antennas requires multivariate and multi-objective optimization processes, which can be accelerated by PSO algorithms. However, the PSO algorithm often fails to achieve optimal results with limited computation resources since the spaces of candidate solutions are quite large for lens antenna designs. This paper presents a design paradigm for multiple-feed lens antennas based on a physics-assisted particle swarm optimization (PA-PSO) algorithm, which guides the swarm of particles based on the laws of physics. As a proof of concept, a design of compact metalens antenna is proposed, which measures a ±55° field of view, a 21 dBi gain with a flatness within 4 dB, a 3-dB bandwidth > 12°, and a compact design with a F-number of 0.2. The proposed PA-PSO algorithm reaches the optimal results 6 times faster than the ordinary PSO algorithm, which shows promising applications on metasurface antenna designs.

**Introduction**

Low Earth orbit satellite communication has the advantages of wide bandwidth, high data rates, and low latency, making low-cost compact antenna designs essential to the ground terminals, such as star dishes.[1] Existing ground terminals are divided into two categories, i.e., servo mechanism antennas and electronically scanned phased array antennas. Servo mechanism antennas are bulky and have a short lifespan, slow scanning speeds, and poor stability for real-time data transmission.[2-4] Electronically scanned phased array antennas exhibit excellent performance, but current electronic scanning methods rely on a large number of phase shifters, resulting in excessively high overall antenna costs.[5-7] Multiple-feed lens antennas can achieve wide-angle beam steering by switching the feeds located on the lens' focal plane, which are promising candidates for low-cost ground terminals of satellite communications.[8-10]

The lens antenna designs have been intensively studied, including geometry lens,[11,12] gradient-index (GRIN) lens,[13-15] and Metalens,[16-20] et. al. The geometric lens often results in a limited field of view (FOV) and a bulky size due to its curved focal plane and large F-number. Although non-planar lenses such as 3D Luneburg lenses can achieve large-angle beam scanning, their feed antennas require conformal shaping, making the antennas less integrated.[21-23] Additionally, non-planar lenses may block the electromagnetic (EM) waves from adjacent antennas when the beam steering angle is large. For example, the FOV of the spherical Luneburg lens antenna is quite limited when forming a tight coupling antenna array. More importantly, non-planar lenses rely on expensive and time-consuming fabrication technologies such as 3D printing, which significantly increases their costs. Planar GRIN lenses highly rely on the choices of materials with low loss tangent and variable permittivity, resulting in high costs and limited functionalities.

Metasurfaces, as two-dimensional artificial electromagnetic materials, tailor the incident wavefront through sub-wavelength unit structures, resulting in compact sizes and light weights.[24-29] More importantly, the metasurfaces offer great design flexibilities to accommodate the incident wavefronts, making them promising candidates for multiple-feed lens antennas. Recently, lens antennas based on metasurfaces, which are named metalens antenna, have been intensively reported. In 2015, Nelson J. G. Fonseca introduced a non-circular symmetrical lens design capable of achieving a 50° scanning angle, but this design required the lens itself to rotate to achieve 360° beam scanning.[30] In 2016, Kien Pham designed a transmissive lens capable of ±30° scanning, yet without a significant breakthrough in scanning angle.[31] In 2019, Hao Fang Wang designed a parabolic-shaped metasurface lens with a scanning angle of ±60°, but the overall gain of the lens was relatively low, with approximately 18.5 dBi at 0° and only around 15 dBi at 60°.[32] Metalens antenna with a compact size and large FOV remains a challenge.

The performances of multiple-feed lens antennas rely on the designs of the lenses, which vary from different functionalities. Firstly, the lens must accommodate different incident wavefronts from different feeds in terms of overall gain and its flatness crossing the FOV. Secondly, the 3-dB angle bandwidth and the number of feeds must be carefully planned to avoid the blind zone within the FOV of the lens antenna. Finally, the cost and compactness of the multiple-feed lens antennas are highly dependent on their lens designs, such as the F-numbers of the lenses, the choices of lens materials, and the fabrication processes of the lenses, et. al. Therefore, the metalens antenna designs are highly dependent on multivariate and multi-objective optimization processes, which can be effectively improved by using PSO algorithms.

PSO is an optimization method that finds the best solution in the search space by moving a population of candidate solutions, named particles, based on a mathematical formula over the particle's position and velocity.[35,36] Therefore, PSO is a good candidate for the optimization of metalens antenna, which has large design flexibilities, i.e. large search spaces. However, the PSO is a metaheuristic procedure, which does not guarantee the optimal solution. In other words, PSO may lead to a sub-optimal solution by directing the swarm of particles towards the closest extreme conditions.

Here, we propose a design paradigm for metalens antenna based on a physics-assisted particle swarm optimization (PA-PSO) algorithm. The solution space of the metalens design has been greatly reduced by applying the rotational symmetry and maximum gain conditions of the metalens phase distribution. As a result, the computation resources for the optimization process are greatly reduced. More importantly, the initial positions of the particles are assigned according to the maximum gain conditions, resulting in less possibility of finding a sub-optimal solution. As a proof-of-concept, we demonstrated a compact and wide FOV metalens antenna design, which measures an FOV of ±55°, F-number of 0.2, a 22 dBi gain with a flatness within 4 dB, and a 3-dB bandwidth > 12°. Compared with the traditional PSO algorithm, the PA-PSO algorithm finds the optimal metalens antenna design using 1/6 computation time, which shows promising applications not only on metalens antenna designs but also on meta-devices design requiring multivariate and multi-objective optimization processes.

**Metalens design based on PA-RSO algorithm**

The proposed multiple-feed metalens antenna consists of a metalens and a feed array as shown in **Figure 1**a and Figure 1b. The metalens is composed of unit cells whose positions can be described by using the polar coordinates $UP(\alpha, r)$, as shown in Figure 1a. The beam steering function of the metalens antenna is realized by switching the feeds. Therefore, the lens antenna emission angle can be described by using the elevation angle $\theta$ and azimuth angle $\phi$ of the deflection beams, i.e., $EA(\theta, \phi)$

The metalens coverts incident wavefronts from different feeds, i.e., Feed 1 … Feed $i$, into planar ones whose deflection angles depend on the location of the feeds, as shown in Figure 1b. Here, the incident waves from Feed $i$ on the metalens can be described by the spatial distribution of the incident electromagnetic waves,

$$E_i(\alpha, r) = A_i e^{i(\omega t + \varphi_i(\alpha, r))} \tag{1}$$

where $A_i$ and $\varphi_i$ are the amplitude and phase of the incident waves, respectively.

The incident waves can be tuned by the unit cells independently in terms of both phase and amplitude. The metalens design has a solution space of the size $M^N$ where $M$ is the set of all possible unit cell selections and $N$ is the number of the metalens unit cells. The solution space of a metalens design is quite large, considering the choice of the unit cells is typically larger than 4 to cover a $2\pi$ phase change and the number of the unit cells is more than 1000, i.e., $M > 4$ and $N > 1000$. In this paper, two assumptions are made to reduce the solution space for the sake of the computation time. One is that the transmission coefficients of the unit cells are unity to minimize the insertion loss of the metalens, i.e. all the unit cells have the same amplitude modulation. The other is that the metalens has rotational symmetry which is the same as the feed array. Therefore, the metalens design can be described by its phase profile, i.e. $\varphi_m(r)$. The output wavefront from the metalens can be written as,

$$E_m = E_i e^{\varphi_m(r)i} \tag{2}$$

The subwavelength unit cell can be considered a point source whose radiation pattern can be decomposed to planar waves with uniform amplitudes. Considering a certain emission angle $EA(\theta,\phi)$, the planar wave component emitted from a unit cell located at $UP(\alpha,r)$ can be expressed as,

$$E_o(\alpha,r) = A_s(\theta,\phi) E_m e^{-\vec{k}\cdot\vec{r}i} \tag{3}$$

where $A_s(\theta,\phi)$ is the weight of the planar wave determined by the scattering properties of the unit cells, $\vec{k}$ is the wavevector of the planar waves, and $\vec{r}$ is the displacement of the unit cell from the center of the metalens. $-\vec{k}\cdot\vec{r}$ is the planar wave phase retardation due to the location of the unit cells. Here, $A_s(\theta,\phi)$ is a constant due to the point source approximation of the unit cells.

Therefore, the EM radiation of the metalens antenna can be expressed as the integration of the planar waves from all the unit cells,

$$E(\theta,\phi) = A_s \int_0^R r \int_0^{2\pi} A_i(\alpha,r) e^{i(\omega t+\varphi_i(\alpha,r)+\varphi_m(r)-\vec{k}\cdot\vec{r})} d\alpha dr \tag{4}$$

where $R$ is the radius of the metalens, $\vec{k}$ is a function of $\theta$ and $\phi$, and $\varphi_i(\alpha,r)$ is determined by the incident wavefront. Therefore, the radiation intensity $I$ of the metalens antenna at a given $\theta$ and $\phi$ can be written as a function of metalens design $\varphi_m(r)$,

$$I(\varphi_m(r)) = E(\varphi_m(r)) \cdot E(\varphi_m(r))^* = [\int_0^R r e^{\varphi_m(r)i} \varepsilon(r) dr] \cdot [\int_0^R r e^{\varphi_m(r)i} \varepsilon(r) dr]^* \tag{5}$$

where $E^*$ is the conjugate of $E$ and $\varepsilon(r) = \int_0^{2\pi} A_s A_i(\alpha,r) e^{i(\omega t+\varphi_i(\alpha,r)-\vec{k}\cdot\vec{r})} d\alpha$ determined by both the incident wavefront and the emission angle of the metalens antenna. Therefore, $I(\varphi_m(r))$ can be written as,

$$I(\varphi_m(r)) = [\int_0^R r e^{[\varphi_m(r)+phase(\varepsilon(r))]i} |\varepsilon(r) dr|] \cdot [\int_0^R r e^{[\varphi_m(r)+phase(\varepsilon(r))]i} |\varepsilon(r) dr|]^* \tag{6}$$

The extrema condition of I can be derived by using the variational method detailed in the supplementary materials,[37] which are

$$\varphi_m(r) + phase(\varepsilon(r)) = \text{constant} \tag{7}$$

or

$$E(\varphi_m(r)) = \int_0^R e^{\varphi_m(r)i}\varepsilon(r)dr = 0 \tag{8}$$

Eq. (7) and Eq. (8) represent the maximum and minimum condition of metalens antenna radiation intensity, respectively.

The extrema condition reveals the relationship between the metalens design $\varphi_m(r)$ and the incident waves from different feeds $\varepsilon(r)$, which are the solution space and the initial condition of the optimization algorithm, respectively. Here, the PSO method is chosen for the metalens antenna optimization algorithm, whose optimization target includes the compactness (F-number), FOV, and the gain. The criterion is to maximize the radiation intensity when F number $\leq 0.2$ and FOV $\geq \pm 50°$.

Figure 1c and Figure 1d show the working principles of the traditional PSO algorithm and the PA-PSO algorithm. The red and blue stars represent optimal and sub-optimal designs, respectively. The red dots and dashed arrows represent the positions and velocities of the particles, respectively. The traditional PSO algorithm guides the swarm of particles using the radiation intensity, while the PA-PSO algorithm guides the swarm of particles based on the extrema condition as shown in Eq. (7). The extrema condition shows the correct directions of the maximum radiation intensity. **Figure S1** shows the flow diagram of the traditional PSO and PA-PSO algorithm, respectively. For a fair comparison, the only difference between the PSO and PA-PSO algorithm is highlighted in the flow diagram with red-dashed lines. The maximum iteration of both algorithms are set to be 5000 to compare the algorithm convergence speed.

**Figure 2**a and Figure 2b show the optimization speed of both algorithms evaluated by using the times of the iteration. Here, the convergence condition is set to be 1% of the radiation intensity variation between adjacent iterations. The PA-PSO algorithm reaches the convergence state after 650 times of iteration while the traditional PSO algorithm costs more than 4000 times of iteration, i.e. more than six times of computation time compared with PA-PSO. Therefore, the PA-PSO approach guides the swarm of particles more efficiently, which reduces not only the computation time but also the likelihood of finding sub-optimal designs.

**Metalens antenna design and characterization results**

The design parameters of the metalens antenna are shown in **Figure 3**a. The feeds array is located on the focal plane of the metalens, which is 22 mm away from the metalens surface, i.e., $F = 22$ mm. The displacement $x$ of the feeds is defined by the distance between the feed and the central axis of the metalens. The radius of the metalens $R$ is 110 mm. The $F$ number of the metalens is defined by the ratio of the focal length $F$ and the lens diameter, i.e., $\frac{F}{2R}$, which is the same as geometric lenses. Therefore, the $F$-number of the proposed metalens is 0.2, indicating a compact design of the metalens antenna.

A Pancharatnam-Berry (PB) unit cell structure is chosen for the metalens design to accommodate the incidence with circular polarization states. [38] The unit cell structure has three layers of identical metal structures spaced by two dielectric layers (Rogers 4350b) with a permittivity of 3.2 and a thickness $h$ of 0.762 mm. The detailed design parameters of the metal layer can be found in the inserted table of Figure 3a. The multiple-layer design of the unit cell structure enables a high transmission and a 2-$\pi$ phase modulation of the incident EM waves with a frequency range of 27 GHz to 30 GHz, which is shown in Figure 3b and Figure 3c, respectively. Here the unit cells are rotated with respect to the fast axis of the left circularized incidence to cover the 2-$\pi$ phase modulation.

A microstrip antenna design is chosen for the feeds of the metalens antenna, as shown in **Figure S2**. The inserted table shows the design parameters. As shown in Figure S2c, the output wavefront of the feed antenna is similar to a point source, which has a 3-dB angular bandwidth of 60° and an axial ratio of less than 3 dB. Figure S2d shows that the insertion loss of the feed is less than 10% within the frequency region, ranging from 27 GHz to 30 GHz.

The metalens antenna is characterized by a microwave near-field scanning system, as shown in **Figure 4**., which is composed of an Agilent N5247A vector network analyzer (VNA), two VNA header extenders at Ka-band (one transmitter and one receiver), and a WR-28 open waveguide as a probe. RF absorbers are also used to avoid reflections between the instrumentation. The lens is placed between the WR-28 waveguide and the feed, which is assembled in the focal plane of the lens. The displacement $x$ of the feed is regulated by a mechanic frame as shown in Figure 4b. As a result, the feed displacement $x$ can be chosen from 0 mm, 5 mm, 10 mm, 15 mm, 20 mm, 25 mm and 30 mm by using different mounting positions on the focal plane. The receiving WR-28 is placed at 100 mm behind the lens to perform the 2D Near-Field measurement as shown in Figure 4d.

**Figure 5** shows the measured gain profiles of the metalens antenna with different feed displacements x when the incident frequency is 28.5 GHz. Figure 5a, Figure 5b and Figure 5c show the comparison between the experimental results (blue lines) and simulation results (red lines) when the displacement is 0 mm, 30 mm and 15 mm, respectively. The experimental results agree well with the simulation results obtained by using the Finite-Difference Time-Domain (FDTD) method. Figure 5d shows the measured gain profiles when the feed displacement varies from 0 mm to 30 mm. The elevation angle is changed from 0° to 55°, indicating a ± 55° FOV of the metalens antenna. The gain reaches 21 dBi when the displacement x is 0 mm, and the flatness of the gain is within 4 dB. The gain profiles of different displacements can be found in the supplementary materials (**Figure S3**). **Table 1** shows a comparison between different types of lens antennas with beam steering function. The proposed metalens design has outstanding performances in terms of compactness and FOV when compared with other metalens antenna designs. The overall performance of the metalens antenna is comparable with the antennas based on 3D dielectric lens but with much lower cost.

**Conclusion**

In summary, this paper proposes a novel PA-PSO optimization method for designing multiple-feed metalens antennas. The proposed method guides the speed and velocity of particles based on the extrema condition of the metalens design, resulting in a much faster optimization process for the metalens antenna design. As a proof of concept, a design of compact metalens antenna is proposed and compared with the existing lens antennas, showing competitive antenna performances in terms of FOV, gain, compactness, et. al., and with a low cost. The proposed PA-PSO algorithm has promising applications in the multivariate and multi-objective optimization processes, including but not limited to metalens antenna designs.

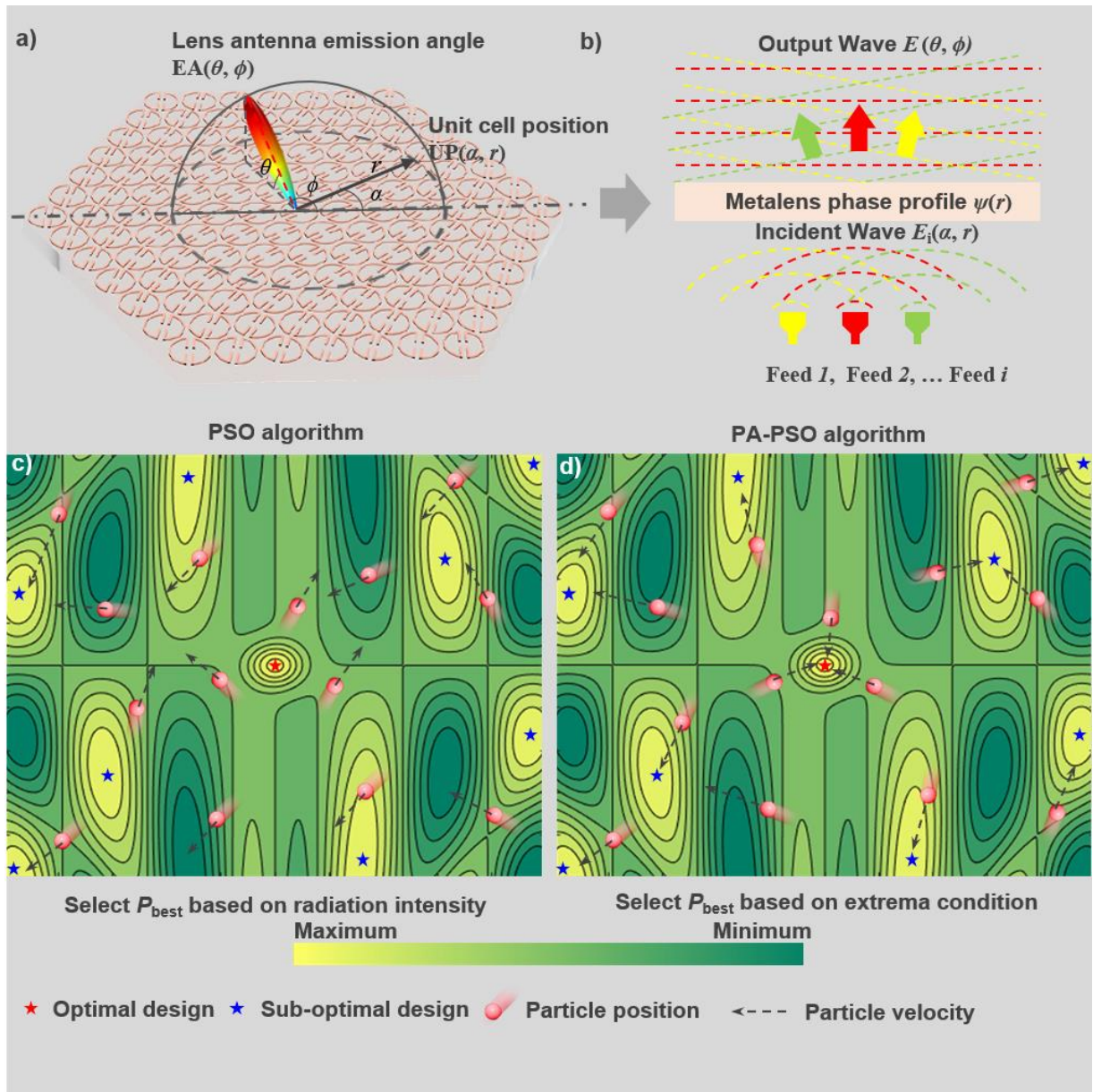

**Figure 1**. Schematics of the PA-RSO algorithm. (a) and (b) the working principle of the metalens antenna. (c) and (d) show the comparison between the traditional PSO and PA-PSO algorithm. The red and blue stars represent optimal and sub-optimal designs, respectively. The red dots and dashed arrows represent the positions and velocities of the particles, respectively. (c) The working principle of PSO algorithm, where the swarm of particles are guided by the radiation intensity. (d) The PA-PSO algorithm guides the swarm of particles based on the extrema condition of the radiation intensity, which shows the correct directions of the maximum radiation intensity. This approach guides the swarm of particles more efficiently, which reduces not only the computation time but also the likelihood of finding sub-optimal designs.

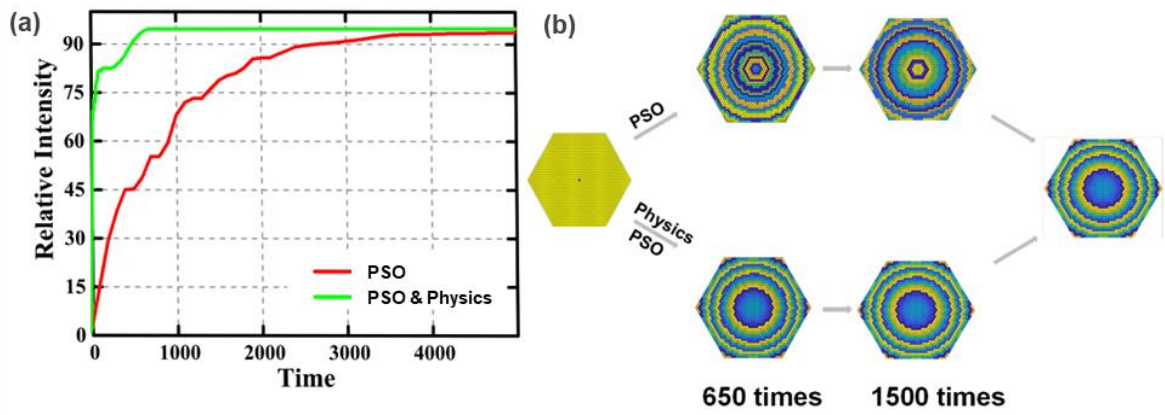

**Figure 2.** (a) Displays the variation of the electric field intensity at a 55° angle with respect to the times of iteration for PA-PSO and PSO algorithms. (b) The phase profile evolution maps for the two algorithms at different times of iteration, respectively.

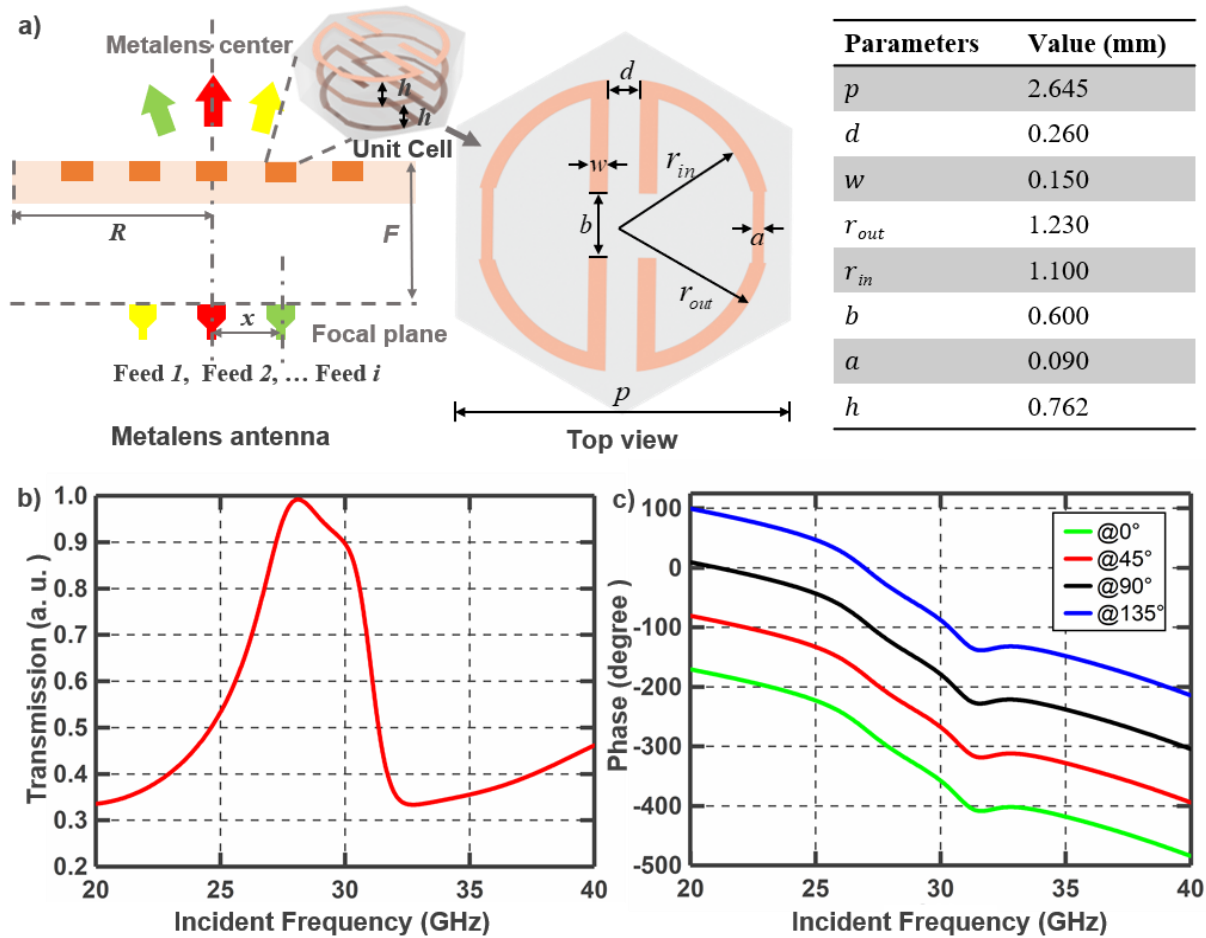

**Figure 3.** (a) The design parameters of the unit structure of the metalens antenna. (b) The transmission curve for the unit cell of the metalens. (c) The phase curve for the unit cell of the metalens, when the unit cell is rotated with respect to the fast axis of the left circularized incidence.

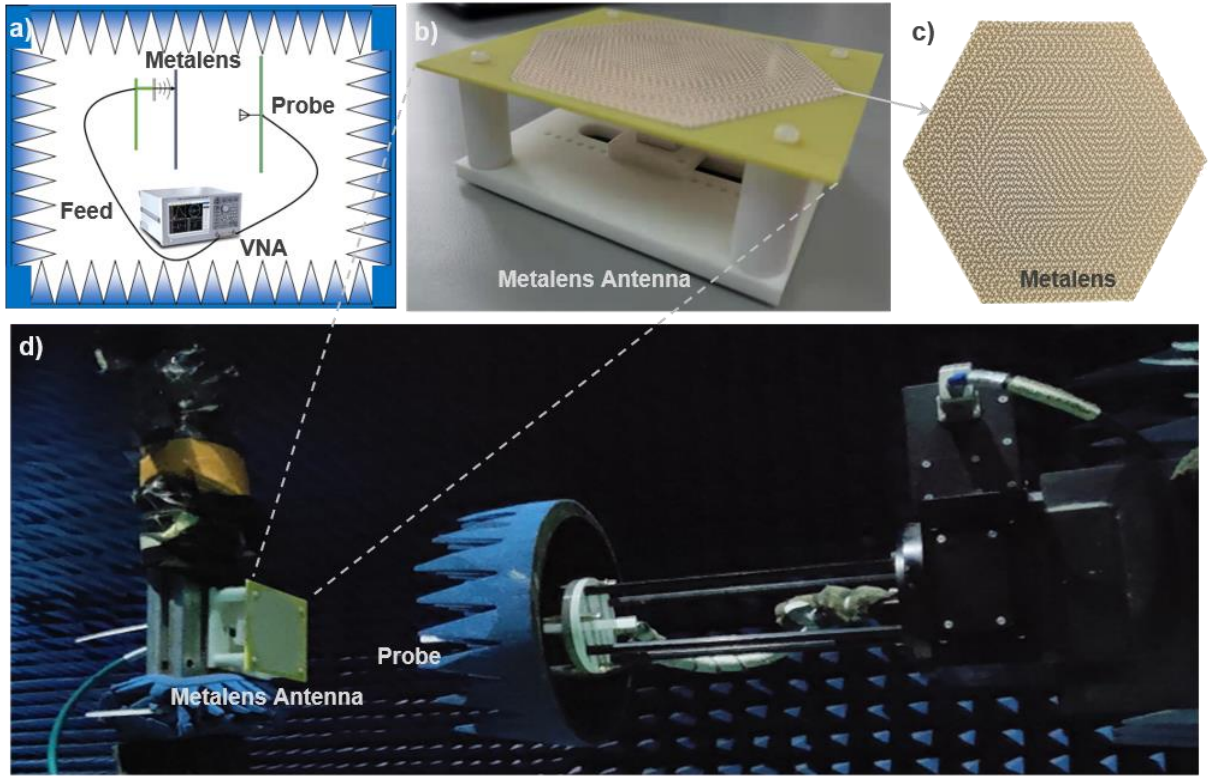

**Figure 4.** Experiment setup and graphs of metalens antenna a) Schematic of the experimental setup for near-field measurement. (b) and (c) Photographs of the metalens antenna and the top view of the metalens, respectively. (d) Photograph of the experimental setup.

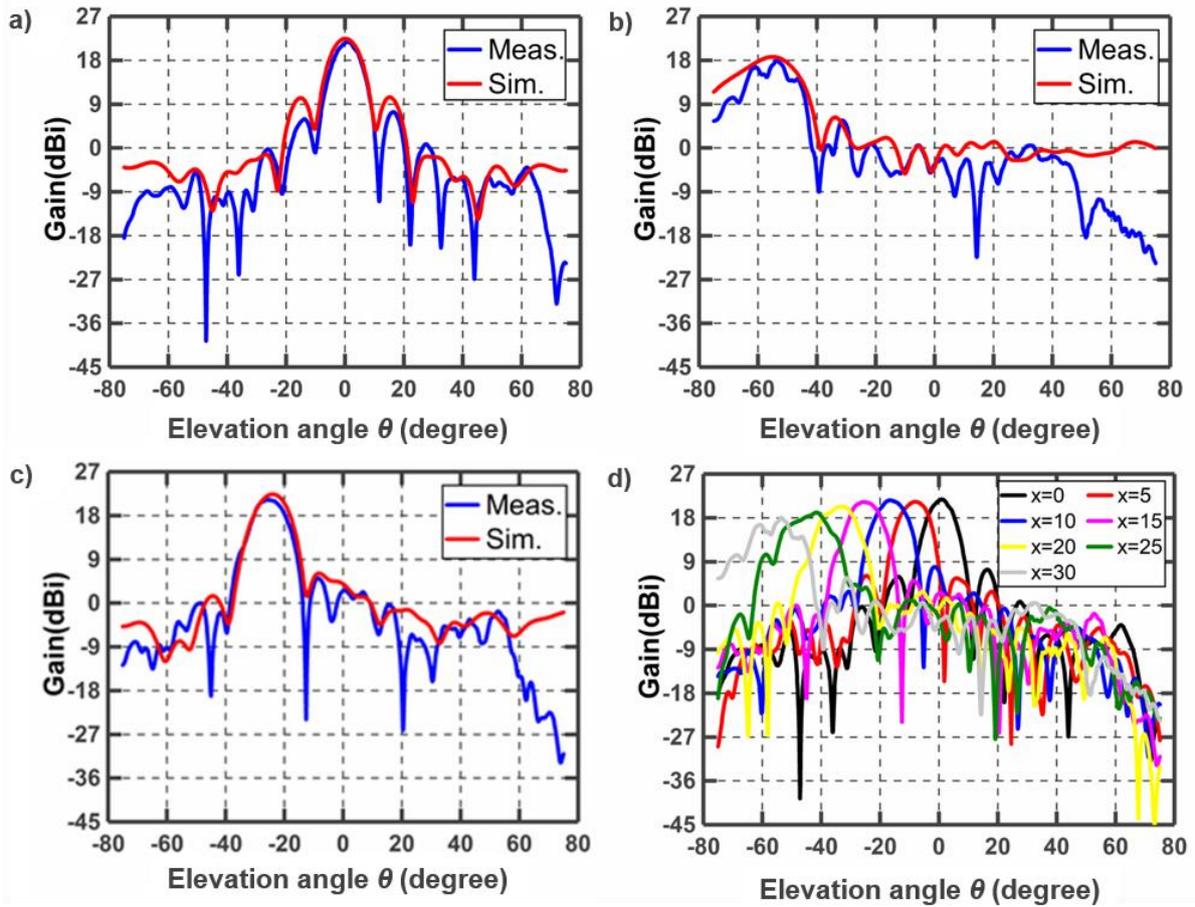

**Figure 5.** The gain profiles of the metalens antenna when the feed is placed on the focal plane with different displacements *x*. (a), (b) and (c) show the comparison between the experimental results (blue lines) and simulation results (red lines) when the displacement is 0 mm, 30 mm and 15 mm, respectively. (d) The measured gain profiles when the feed displacement varies from 0 mm to 30 mm. The elevation angle is changed from 0° to 55°, indicating a ± 55° FOV of the metalens antenna

Table 1: The comparison between different types of lens antennas with beam steering function.

| Lens Type | Scan Angle | 3dB Beamwidth | Gain (dBi) | Band (GHz) | Size (mm) | F | Steering Method | Cost | Ref. |
|---|---|---|---|---|---|---|---|---|---|
| Luneburg Lens | ±55° | 15.7° | 19 | 26-40 | D=60 H=60 | NA | Nonplanar Feed Array | High | [22] |
| | ±70° | 12.8° | 21.2 | 36.5-38 | D=50 H=50 | NA | | High | [21] |
| Grin Lens | ±48° | 4° | 22.8 | 26.5-40 | D=152.4 H=112 | 0.5 | Planar Feed Array | High | [13] |
| | ±40° | 17.1° | 24 | 27-29 | D=80 H=35.1 | 0.4 | | High | [14] |
| Metalens | ±55° | 12° | 21 | 27-30 | D=110 H=1.524 | 0.2 | | low | This work |
| | -30°-60° | NA | 27.5 | 20-30 | 192.5*143.5 H=9.7 | <1 | Feed Array + Rotation | Medium | [32] |
| | 0-50° | 6.48° | 27.3 | 29.5-30 | 195*145 H=3.14 | 0.55 | | Medium | [33] |

**Acknowledgements**

This work was supported by the National Natural Science Foundation of China (Grant Nos. 61975026 and 61875030) and Creative Research Groups of the National Natural Science Foundation of Sichuan Province (2023NSFSC1973). Shibin Jiang and Wenjun Deng contributed equally to this work.